# Adaptation and learning of molecular networks as a description of cancer development at the systems-level: potential use in anti-cancer therapies

Dávid M. Gyurkó[1], Dániel V. Veres[1], Dezső Módos[1,2,3], Katalin Lenti[2], Tamás Korcsmáros[3] and Peter Csermely[1,1]

[1]*Semmelweis University, Department of Medical Chemistry, Tuzolto u. 37-47, H-1094 Budapest, Hungary;* [2]*Semmelweis University, Department of Morphology and Physiology, Faculty of Health Sciences, Vas u. 17, H-1088, Budapest, Hungary;* [3]*Eötvös Loránd University, Department of Genetics, Pazmany P. s. 1c, H-1117 Budapest, Hungary*

**Abstract**: There is a widening recognition that cancer cells are products of complex developmental processes. Carcinogenesis and metastasis formation are increasingly described as systems-level, network phenomena. Here we propose that malignant transformation is a two-phase process, where an initial increase of system plasticity is followed by a decrease of plasticity at late stages of carcinogenesis as a model of cellular learning. We describe the hallmarks of increased system plasticity of early, tumor initiating cells, such as increased noise, entropy, conformational and phenotypic plasticity, physical deformability, cell heterogeneity and network rearrangements. Finally, we argue that the large structural changes of molecular networks during cancer development necessitate a rather different targeting strategy in early and late phase of carcinogenesis. Plastic networks of early phase cancer development need a central hit, while rigid networks of late stage primary tumors or established metastases should be attacked by the network influence strategy, such as by edgetic, multi-target, or allo-network drugs. Cancer stem cells need special diagnosis and targeting, since their dormant and rapidly proliferating forms may have more rigid, or more plastic networks, respectively. The extremely high ability to change their rigidity/plasticity may be a key differentiating hallmark of cancer stem cells. The application of early stage-optimized anti-cancer drugs to late-stage patients may be a reason of many failures in anti-cancer therapies. Our hypotheses presented here underlie the need for patient-specific multi-target therapies applying the correct ratio of central hits and network influences — in an optimal sequence.

In this paper first we will describe cancer development as a two-phase phenomenon characterized by a first increased than decreased plasticity (or in alternative wording: by a first decreasing than increasing rigidity) at the systems-level. We will propose that cancer stem cells have the unique property to induce rapid and gross changes of their network plasticity/rigidity. Next, we will list the cancer-specific properties of molecular networks highlighting the adaptation of network structure and dynamics in various stages of cancer development as a model of cellular learning. Finally, we will highlight that different network-related anti-cancer strategies are needed to target early and late stage cancer cells, as well as cancer stem cells, and will list the possibilities to target all these cell communities.

## 1. Development of cancer as a learning process of increasing and then decreasing plasticity at the systems-level

Cancer cells are products of a complex cell transformation process. The starting steps of this process are often mutations or DNA-rearrangements, which destabilize the former cellular phenotype. As a result, a cell population with a large variability in chromatin organization (including DNA-hypomethylation, histone modification and chromatin structure), gene expression patterns and interactome composition is formed. This heterogeneous cell population is characterized by an increased level of stochastic processes (noise), phenotypic plasticity [1–6], and by an increase in the network entropy of protein-protein interaction networks (the definitions of local and more global variants of network entropy see in [7–9]). Higher degree-entropy of signaling networks was found to correlate with lower survival of prostate cancer patients [9], which underlies both the medical relevance and the generality of these changes. Increased heterogeneity may help survival not only at the level of individual cells, but also by allowing the formation of cooperating inter-cellular networks of cancer cells [10,11]. This cooperative, paracrine cross-regulation may change to a selfish, autocrine development later [12,13].

---

[1]Corresponding author; Csermely.Peter@med.semmelweis-univ.hu

In many general aspects, the development of cancer cells resembles to that resulting in changes of cellular phenotypes e.g. during embryogenesis. Major steps of the epithelial-mesenchymal transition (where epithelial cells lose both their apical/basal polarity and original cellular contacts, gain the ability to transverse the extracellular matrix, and ultimately contribute to tissues other than the original epithelial sheet) correspond well to the major stages of metastasis formation [14,15].

Here we argue that malignant transformation is essentially a two-phase process. We propose that the first phase of malignant transformation is characterized by an increase of plasticity in the protein-protein interaction, signaling and other networks of transformed cells, which is followed by decrease of network plasticity (Fig. 1), where network plasticity (or in other words network flexibility) is related to the degrees of freedom of network nodes [16]. The initial stage may correspond to "clonal expansion", and the appearance of tumor initiating cells. The late stage may represent either late-stage primary tumor cells, or metastatic cells, which already settled in their novel tissue environment.

Cancer stem cells [17] can possess both types of the above duality. Cancer stem cells can display a dormant phenotype, which is not rapidly proliferating, and has a more rigid structure of its protein-protein interaction, signaling and other networks than that of rapidly proliferating early stage tumor cells. Importantly, cancer stem cells may also possess a rapidly proliferating phenotype [17], which may have a highly plastic network structure. Therefore, cancer stem cells are not discriminated by the plasticity of their actual network structures (called as structural plasticity [16]), but by their especially high ability to modulate the plasticity of their networks (called as dynamic plasticity [16]) according to the needs of the environment. It is important to note that the specially high ability of cancer stem cells to modulate the plasticity of their networks is in accordance with the original cancer stem cell definition of tumor initiation after serial dilutions [17]. A highly increased ability of plasticity modulation (which results in an increased level of evolvability) may prove to be a major discriminatory hallmark of cancer stem cells. Importantly, this increased plasticity modulation ability may be a key reason why anti-cancer therapies often induce cancer stem cells instead of killing or transforming them.

There are many signs of the initial increase of system plasticity during cancer development, such as the increased heterogeneity, noise and entropy mentioned before. Tumor initiating cells showed a larger plasticity also at the level of physical deformability [18,19]. However, at late stage carcinogenesis of the primary tumor, or when tumor cells already established metastases and were incorporated to a more stable tissue environment, the systems-level plasticity of cancer cells may decrease again.

The cancer-specific, metastable states, which were termed as "cancer attractors" by Stuart Kauffman in 1971 [3,20] may be especially typical to these later stages of cancer development. Additionally, early stages of cancer development may be characterized by numerous "shallow" cancer attractors developing a more plastic structural network of the cell residing in transit on this relatively smooth state space environment, while late stages of tumor development may involve fewer but "deeper" cancer attractors, where the cancer cell becomes stabilized in this more rough state space environment. The dual change described above corresponds well to various steps in the transition to cancer attractors, since cancer cells should first cross a barrier in the (epigenetic) fitness landscape, which might be lowered by mutations or epigenetic changes [3], but still requires a transient destabilization of the transforming cell, which leads to a more plastic phenotype with all the phenotypic characteristics described before. The phenotype of the already established, late-stage cancer cells is still more plastic and immature than that of normal cells, but may often be more rigid than the phenotype of the cells in the intermediate stages of carcinogenesis (Fig. 1).

Such a biphasic change resembles to that of cell differentiation processes, where an initial increase of entropy of chromosomal order and co-regulated gene expression pattern is followed by a later decrease [21]. An analogous set of events happens in cellular reprogramming, where single-cell studies revealed that an early stage, very heterogeneous, stochastic phase is followed by a late phase, which is programmed by a hierarchical set of transcription factors [22]. The recently discovered super-enhancers [23,24] may characterize this second, consolidated phase of phenotype rearrangement. Importantly, a more ordered system is generally less controllable than a disordered one [25], which warns that (A) therapeutic interventions of early stage of carcinogenesis are more efficient than those against late-stage tumors; (B) late-stage tumors should be attacked by a fully different strategy than early-stage tumors preferring indirect targets having a smaller centrality in molecular networks, which cause less side effects and toxicity in these more rigid, late-stage cancer-specific networks/cellular systems, see Fig. 1 and [26].

## 2. Network segments participating in adaptive processes

The cancer development process is increasingly described as a systems-level, network phenomenon [27,28]. Applying this description to the cancer development stages described above, in this section we summarize the adaptation options of network structures in general. In Section 3, we will describe the cancer-specific network adaptation events.

Adaptation of networks is often described as "network evolution", where the term refers to changes in the network contact structure (e.g. changes of parameters like network connectivity, edge weights, diameter, centrality, motifs or modules [29,30]). The identification of these changes has a predictive potential both in retrospect and about future development of the complex system represented by the network. Network evolution may follow different timescales varying from microseconds to decades or more [31–33]. Thus the assessment of molecular network changes in the progression of cancer requires a very careful selection of timeframes both in clinical sampling and in network analysis.

Network dynamics extends the frame of network evolution to the refined changes of network nodes (e.g. individual proteins) when transmitting signals, or participating in their cellular function requiring a concerted action of multiple proteins. Network dynamics is strongly related to the underlying network topology. Transitions to a new state, as it was seen in tumor development [34], have key importance in network dynamics.

Hubs (i.e. nodes with much more neighbors than the average) are key determinants of local network topology [35]. A key example of these highly connected hubs is the p53 protein. The p53 tumor suppressor is indeed a key regulator of cancer-related molecular networks [36–38]. Many different pathways may converge on hubs, which are often evolutionary conserved proteins [39]. Reversely, a hub often acts as a distributor of perturbations. Network motifs are repeated local patterns of network topology typically consisting of 3 to 6 nodes. Negative feedback loops and feed-forward loops are significant motifs in adaptation [40,41].

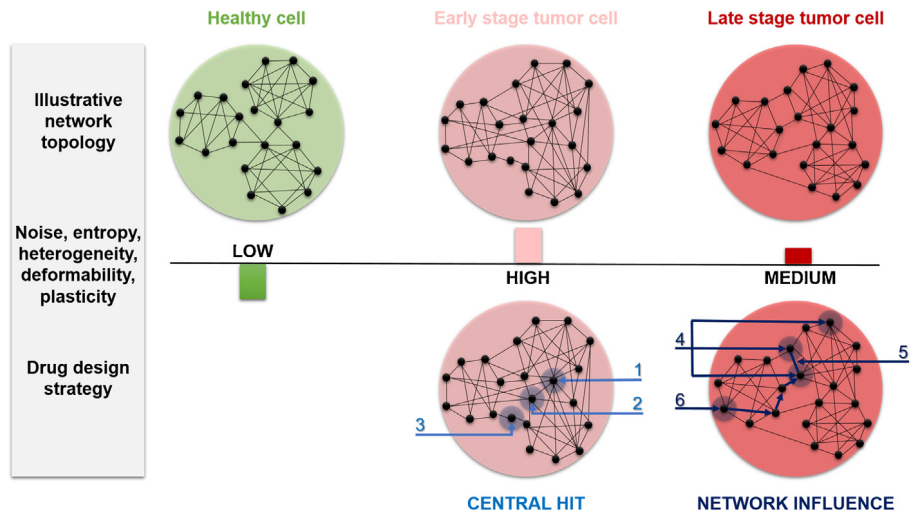

**Fig. 1.** Development of cancer as a learning process of increasing and then decreasing plasticity at the systems-level determines the most appropriate drug targeting strategy. The figure illustrates the major hypothesis of the current review showing that the transition of cells to a cancer attractor during malignant transformation is essentially a two-phase process. We propose that in this transformation an initial increase of system plasticity is followed by a late-stage decrease of plasticity. Plastic, early stage systems of cancer development are characterized by a higher level noise, entropy and physical deformability. The plastic → rigid transition of network structure during cancer development invokes a rather different targeting strategy in early and late phase of carcinogenesis. While at the early phase of cancer development hitting of central nodes (such as hubs, marked as "1"; inter-modular bridges, marked as "2"; or bottlenecks, marked as "3") being in the cross-roads of regulatory processes may be a winning strategy, at later stages of carcinogenesis the more indirect means of network influence strategy, such as the multi-target (marked as "4"), edgetic (marked as "5") or allo-network drugs (marked as "6") [26,79,80] should be used. Importantly, heterogeneous cancer cell populations [6,92] may harbor early- and late-stage tumor cells at the same time. Moreover, cancer stem cells [16] may have the ability to change their "plasticity phenotype" from that of early- to late-stage tumor cells and vice versa. Therefore, multi-target, combinatorial or sequential therapies using both central hit- and network influence-type drugs may provide an even more useful anti-cancer therapeutic modality than previously thought.

Network modules are groups of nodes connected more densely to each other than to the rest of the network. In molecular networks a spatial or temporal module often means a functional unit [42]. Altered modularity was shown as a predictive measure in breast cancer prognosis [43]. Most modules of cellular networks have a high overlap with each other [44,45]. Modules of the yeast protein–protein interaction network are becoming more condensed and displaying smaller overlaps upon stress [46]. In extreme cases stress may result in the disintegration of the network leading to the death of the organism, which may be an important part of the network-level mechanism of action of several anti-cancer drugs.

Modules may segregate network segments, which are more plastic and/or more rigid. While rigid network segments preserve the result of a past adaptive process often displaying an optimized function, flexible network segments are capable of plastic adaptation to present or future challenges of the environment. While rigid network segments are able for fast signal transmission without a large dissipation, plastic network segments have a slower signal transmission and a larger dissipation [16,26,47].

Bridges are node-pairs connecting modules, and a bottleneck is a key inter-modular node. Network perturbations have to propagate through these nodes, therefore bridges and bottlenecks are important points of regulation and adaptation [48,49]. The so-called creative nodes [50] are joining multiple modules in a highly dynamic fashion. Creative nodes connect functionally distinct modules. Therefore, the abundance of these exceptionally unpredictable nodes may be a key regulator of "adaptation-speed" and related system plasticity and evolvability. Fig. 2 illustrates the most important structural elements of network adaptation.

## 3. Cancer-specific properties of molecular networks

After the description of key network segments participating in adaptive processes in general, here we summarize the current knowledge on the most important changes of molecular networks in cancer development. Table 1 highlights a few proteins playing a key role in cancer-related molecular networks.

Protein structure networks have nodes of proteins as amino acids, where the edge weight depends on the distance of two neighboring amino acids in the 3D structure of the protein. Detailed studies on the network representation of cancer-specific proteins and their mutations are missing. However, recurrent findings showed that intrinsically disordered proteins play an important role in cancer-specific cellular events [51–53]. Additionally, cancer-related proteins have smaller, more planar, more charged and less hydrophobic binding interfaces than other proteins, which may indicate a low affinity and high specificity of cancer-related interactions [54]. An increased "conformational noise" in cancer represented by intrinsically disordered proteins and by low affinity interactions would be in agreement with the more plastic network structure of tumor initiating cells. Our hypothesis would predict a decreased expression of disordered proteins or proteins with low affinity interactions in the late-phase of cancer development (i.e. in late stage primary tumors, or when metastatic cells already settled in the novel tissue environment) as opposed to the early phases of "clonal expansion", or cancer stem cell formation, where the expression of these proteins may be higher.

Protein–protein interaction networks (interactomes) have the individual proteins (or their domains) as nodes, and their physical interactions as edges. Cancer-specific interactomes may be constructed by taking into account protein abundances, or (as a first approximation) incorporating mRNA expression patterns [48,55,56]. A microarray gene profiling study showed that genes with elevated expression are coding well-connected proteins, while suppressed genes code less connected proteins in lung squamous cancer [57]. Interactome-based studies indicated that proteins with cancer-specific mutations are hubs with a tendency to form a "rich-club", i.e. being connected with each other [58], and are located in the center of modules or create bridges between them [59]. Inter-modular hubs, such as IRS1 or BRAF, have a particularly important role in oncogenesis [43]. Cancer-related proteins tend to form transient interactions and usually involved in multiple pathways [54]. Dynamical rearrangements of interactome segments may contribute to the changes in plasticity during cancer development. Gene-expression enriched protein–protein interaction networks of 6 cancer types showed an increased network entropy, if compared to interactomes of healthy cells [8]. Increased network-entropy may be an important ingredient of the increased plasticity of malignant cells, and may significantly contribute to the increased robustness against stress and other environmental stimuli.

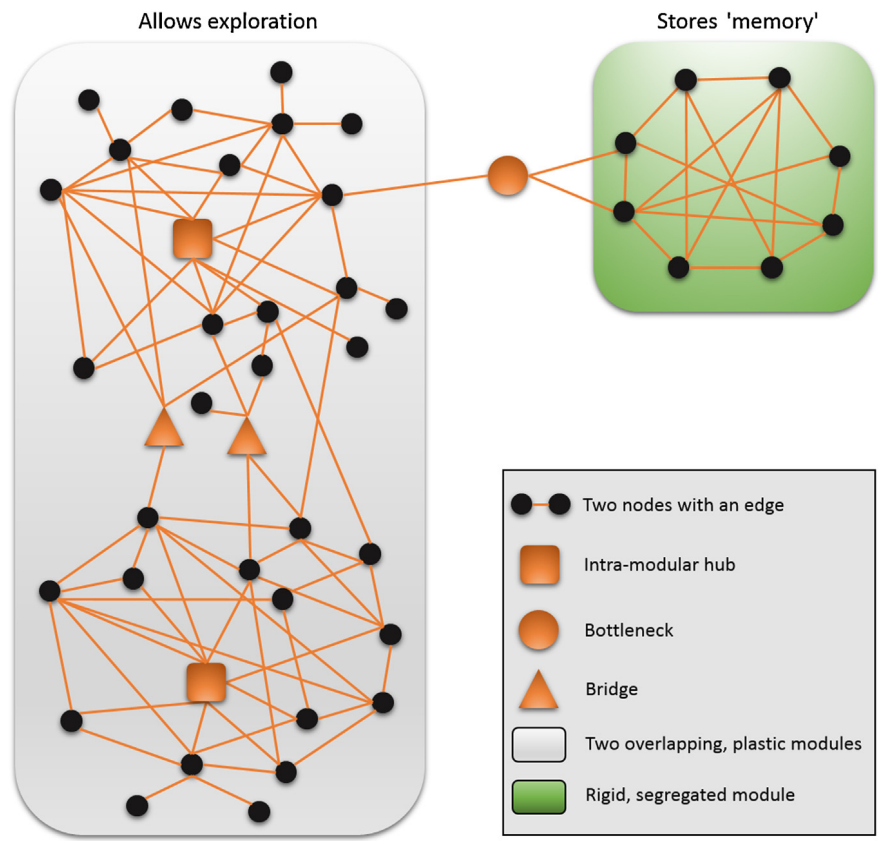

**Fig. 2.** Examples of key structural segments participating in network adaptation. The figure illustrates a few key structural network segments, which often play a key role in the adaptation of complex systems. Intra-modular hubs are highly connected, central nodes collecting and distributing network perturbations. Bridges and bottlenecks are crucial regulators of information flow between network modules. Network modules often functional units and may display a rigid or flexible structure. While the former has been optimized to a certain function, and may serve as a "memory" of the network preserving the result of successful adaptation to past events, the latter may provide the plasticity promoting adaptive processes in the present or future.

The autonomy of the cancer cell is based mainly on the changes in its signaling network, formed by interconnected signaling pathways including its gene regulatory network [60]. Changes in the expression level of signaling proteins in cancer cells may cause the activation of major cancer-related pathways, and can rewire the whole signaling network, as it was shown in the case of the epidermal growth factor receptor (ERBB1) signaling network [61]. Cross-talks between the individual signaling pathways are able to create a high number of new input-output combinations increasing the plasticity of the signaling network [2]. Earlier, we found a significant change in the expression level of specific cross-talking proteins in hepatocellular carcinoma leading to a highly different, malignant network of signaling pathways [62]. The up-regulation and cross-talks of RAS-ERK and PI3K-mTORC1 pathways play a key role in the progression of several tumor types [63]. The regulation of signaling cross-talks highly depends on the spatial organization of the cell involving differences in molecular crowding [64], scaffold proteins [65], protein translocation [66] or membrane association [67]. Increased network entropy was found in the signaling network of the metastasizing phenotype of breast cancer, if compared to non-metastasizing cells [7]. This may indicate the increased plasticity of cancer-specific signaling in the early phases of tumor development. Our recent studies showed that increased plasticity of signaling networks characterizes the initial stages of tumor development (such as adenomas) better than later stages of malignant transformation (such as carcinomas; Dezső Módos, Katalin Lenti, Tamás Korcsmáros and Péter Csermely, in preparation).

**Table 1**
Network drug target options.

| Molecular network type | Possible drug targets | Cancer-specific network properties of possible drug targets | References |
|---|---|---|---|
| Protein–protein interaction network | p53, CDK6 | Hub inactivation, bottleneck dysfunction | [36,49] |
| Signaling network | ERBB1 | Edgetic perturbation of receptor tyrosine kinase pathway | [61] |
| | PDGFR, TGFBR | Cross-talks of different pathways during epithelial-mesenchymal transition | [95] |
| Genetic interaction network | MYC, NES, MMP2, FOS, TNC | Hub over-activation, hub-driving node under-activation | [96] |
| Chromatin network | BRD4 | Super-enhancer participating hub | [23] |
| Metabolic network | Fumarate hydratase, Succinate dehydrogenase, PKM2 | Targeting the Warburg effect, Increased metabolic flux | [97] [98] |

Transcriptional regulation is often executed by a hierarchical cascade of transcriptional factors. The recently discovered super-enhancers [23,24] showed that master transcription factors may often establish a cancer-specific super-enhancer locus. Whether the organization of these super-enhancer regions represents a late-phase of carcinogenesis, and is preceded by a very heterogeneous, stochastic early phase, like that in cell reprogramming [22], is an open question of future studies.

Chromatin networks are formed by connections between distant DNA-segments, and can be used for the explanation of the chromosomal alterations in cancer [68]. The overexpression of the prostate cancer oncogenic transcription factor, ERG induced a profound effect on chromatin network configuration [69]. Further studies are needed to assess the cancer-specific alterations of transcription-related chromatin networks, such as that describing the connection pattern established by the active RNA-polymerase II-complex [70].

Regulation or mutations of microRNAs play a crucial role in the regulation of cancer-specific gene expression [71]. MicroRNAs can function either as tumor suppressors, such as the let-7 family members targeting the RAS signaling pathway in breast cancer [72], or as oncogenes, like the miR-155 targeting key breast cancer genes, such as FOXO3A or RHOA [73]. The analyses of cancer-specific microRNA networks showed more disjointed subnetworks than in normal tissues [74]. Disjoint modules decrease the interdependence of modular changes, and also serve as an adaptive response in stress [46]. Disjoint modules confer a greater plasticity to the whole system.

Metabolic networks consist of major metabolites as nodes connected by specific enzymatic reactions converting them to each other [75]. Global changes in metabolic networks contribute to the Warburg-effect, and are driven by key enzymes such as the PKM2 isoenzyme of pyruvate kinase [76,77].

The stage-specific changes of molecular networks in tumorigenesis are still largely unexplored, because we usually have detailed systems-level network data only on the two endpoints, i.e. the healthy tissue and the developed late tumors. Therefore, the assessment of consecutive changes in network structure and dynamics during intermediate, consecutive steps of malignant cell and metastasis formation are key steps of future systems-level analysis of cancer development. Network analysis of cancer stem cells will provide a special understanding of both plasticity modulation and evolvability at the systems-level.

## 4. Network-related anti-cancer therapies

It is a growing challenge to identify new drug targets and efficient combination of drugs in order to design new anti-cancer therapies with less toxicity, side-effects and resistance development. As we have shown in the preceding section molecular networks at various levels greatly improved our systems-level understanding of tumor initiation and progression [26–28,78].

Recently a dual network strategy, the central hit strategy and the network influence strategy, was described to target various diseases [26]. In cancer both strategies may be used. Using the central hit strategy our aim is to damage the network integrity of the malignant cell in a selective manner. This may be the most important strategy to attack tumor initiating cells at their first stage of development (e.g. early stages of malignant transformation of metastasis formation). Using the network influence strategy we would like to shift back the malfunctioning network to its normal state. This may be the most important way to shift back tumor cells from their cancer attractors reached in late stages of their development (e.g. late stage primary tumors or after the metastatic cells settled to their new tissue environment). The central hit strategy is applied against a population of highly destabilized cells having a very plastic network structure. Efficient targeting of these systems requires the targeting of their central nodes/edges. In contrast, the network influence strategy is applied to cells, which have a much more rigid network structure than the networks of highly undifferentiated or dedifferentiated cells. Targeting the central nodes/edges of systems having a low plasticity may easily 'over-saturate' the system causing side-effects and toxicity. Therefore, the network influence strategy often needs an indirect approach, where e.g. neighbors of the real target are targeted (allo-network drugs [79]). In another indirect approach of the network influence strategy multiple targets are targeted at the same time 'mildly' (using multi-target drugs [80]), and their indirect and/or superposing effects lead to the reconfiguration of diseased network state back to normal.

The central hit strategy often uses the key central position of inter-modular hubs. These central nodes are often important oncogenes serving as targets in drug development [43]. Similarly, in signaling networks protein or microRNA hubs [48,57,58,81,82], inter-modular positions of cross-talks [83], bottlenecks [84] as well as multi-pathway proteins offer important target candidates [43,62] as central hits.

However, the central hit strategy may often attack nodes, which are so central that their inhibition damages key functions of healthy cells. As an example, mTOR, the mammalian target of rapamycin is an important multi-pathway protein, which mutates in most of the tumors and thus causes hyper-active phenotype [85]. Because of the super-central position of mTOR, edgetic drugs targeting single mTOR interactions have a much more selective effect [86] than targeting all mTOR interactions by conventional drugs [87]. Another example of edgetic anti-cancer therapies is the inhibition of p53/MDM2 connection by nutlins, which liberate the tumor-suppressor effect of p53 from MDM2-induced inhibition [88].

The network influence strategy often implies multi-target attacks [80]. These attacks may allow the simultaneous targeting of two or more peripheral nodes instead of targeting a single, central node [25]. Such combinatorial or multi-target therapies can be identified using interactomes, signaling or metabolic networks [89,90] and led to the design a simultaneous targeting therapy e.g. in colorectal cancer [91]. The systematic development of allo-network drugs [79] acting at the neighborhood of real targets and specifically transmitting their signals to them requires a more detailed knowledge of both indirect targeting and the molecular details of allosteric action, and therefore remains and major challenge of future studies.

## 5. Network adaptation in various stages of cancer development as a key aspect of anti-cancer therapies

Here we propose that the large structural changes of molecular networks during cancer development require a rather different targeting strategy in early and late phase of carcinogenesis (Fig. 1). While at the early phase of cancer development hitting of central nodes in various molecular networks (such as inter-modular hubs being in the cross-roads of regulatory processes) may become a winning strategy, the very same pharmacological intervention may become useless at a later stage of carcinogenesis, where the primary tumor was already established, or the metastasized cells have been settled in their new tissue environment. These late-stage tumor cells should be attacked by the more indirect means of network influence strategy, such as by edgetic, multi-target, or allo-network drugs [26,79,80].

The analysis of Rajapakse et al. [25] offers important clues on targeting molecular networks during cancer development. Referring to the general observation that symmetrical networks are more uncontrollable, they hypothesize that the highly heterogeneous, highly plastic, intermediate state of tumor-initiating cells right before or at attaining their metastatic potential is the most difficult to attack by well-targeted external inputs. This is the phase, where only the central-hit strategy may be applied. However, at late stage primary or metastatic tumors, where the network structures became more asymmetric again (Dezső Módos, Katalin Lenti, Tamás Korcsmáros and Péter Csermely, in preparation) only the indirect effects of the network influence strategy may be successful.

Very importantly, cancers often harbor cancer stem cells, which may be induced by conventional anti-cancer therapies, themselves [16]. These stem cell-like tumor cell subpopulations possess an extraordinarily high ability to change their plasticity. When cancer stem cells acquire a more dormant state, they may change their networks from a more plastic to a more rigid structure. On the contrary, cancer stem cells may have a transition from a more rigid to a more plastic network structure, if they form early tumor progenitor cells. This large level of dynamic plasticity may be a major reason of the development of drug resistance in many cancer cases.

The presence of highly heterogeneous cell populations (including various forms of cancer stem cells) in a cancer patient, as well as the extreme heterogeneity of patient subtypes [6,92] all give a particular importance of sequential multi-target therapies, where multiple drug treatments are given in a particular order using a well-designed temporal pattern of consecutive treatments. In agreement with Kitano [1] and many key results in the field (e.g. when epidermal growth factor receptor inhibitors sensitized cancer cells to subsequent DNA-damage by unmasking an apoptotic pathway in breast cancer see the work of [93] and the review of Huang and Kauffman in this issue) we strongly believe that sequential multi-target therapy will be the major mode of intervention in cancer therapies of the future.

## 6. Conclusions and perspectives

In conclusion, in this review we proposed that malignant transformation is a two-phase process, where an initial increase of system plasticity is followed by a decrease of plasticity at late stages of carcinogenesis of primary tumor cells or of metastatic cells already settled in their novel tissue environment. The increased systems-level plasticity of cancer initiating cells is characterized by an increased

- level of stochastic processes (noise);
- network entropy;
- "conformational noise" represented by intrinsically disordered proteins and by low affinity interactions;
- phenotypic plasticity;
- cell heterogeneity;
- physical deformability and
- dissociation of network modules (Fig. 1).

Our hypothesis would predict a decrease in all the above parameters in the late-phases of cancer development as opposed to the early phases described above. In late phase carcinogenesis the increase of system rigidity may be accompanied by an increased hierarchy of regulatory processes.

The large structural changes of molecular networks during cancer development require a rather different targeting strategy in early and late phase of carcinogenesis. Importantly, a more ordered system is generally less controllable than a disordered one [25], which warns that (A) therapeutic interventions of the early, more plastic stage of carcinogenesis are more efficient than those against late-stage tumors; (B) late-stage tumors should be attacked by an entirely different strategy than early-stage tumors, see Fig. 1 and [26]. Besides other therapeutic modalities (such as surgery, radiotherapy etc.) at the early phase of cancer development hitting of central nodes of cancerous molecular networks (such as inter-modular hubs being in the cross-roads of regulatory processes) may become a winning strategy killing the very heterogeneous cancer cell population. However, central hit-type pharmacological interventions may become useless at a later stage of carcinogenesis. Late stage primary tumors or the already metastasized cells in their new tissue environment should be attacked by the more indirect means of network influence strategy, such as by edgetic, multi-target, or allo-network drugs see Fig. 1 and [26,79,80], since here not the eradication of a large heterogeneity of tumor cells, but the shift of the malignant cellular network to a less malignant state is the desired action.

Regretfully, many in vitro test systems of anti-cancer drug candidates resemble to the plastic cellular systems of early stage cancer development, while disease is usually detected in patients, when it reached the late phase of development [94]. The application of early stage-optimized anti-cancer drugs to late-stage patients may be a reason of many failures in anti-cancer therapies. This situation underlies the importance of systems- level studies of cancer cell development and the network analysis of the data obtained.

Importantly, the initial increase and later decrease of system plasticity during cancer development may appear both at the level of the network of individual cells and at the level of the network of cell populations. Heterogeneous cancer cell populations may harbor early- and late-stage tumor cells at the same time. Moreover, cancer stem cells may have the ability to change their "plasticity phenotype" from that of early- to late-stage tumor cells and vice versa. Therefore, multi-target, combinatorial or sequential therapies using both central hit- and network influence-type drugs may provide an even more useful anti-cancer therapeutic modality than previously thought.

## Conflict of interest

Authors declare no conflict of interest.

## Acknowledgements

Work in the authors' laboratory was supported by research grants from the Hungarian National Science Foundation (OTKA-K83314), from the EU (TÁMOP-4.2.2/B-10/1-2010-0013) and a János Bolyai Scholarship to TK.